\documentclass[aps,amssymb,amsmath, twocolumn, pra, superscriptaddress]{revtex4}
\usepackage{graphicx}
\usepackage{epsfig}
\usepackage{amsmath}
\usepackage{bm}
\usepackage{braket}
\usepackage{natbib}
\usepackage{enumitem}
\usepackage{longtable}
\graphicspath{ {/home/asarkar/Pictures/} }
\usepackage{dsfont}

\usepackage[colorlinks, citecolor= magenta]{hyperref}
\usepackage[up]{subfigure}

\usepackage{epstopdf}

\usepackage{booktabs}

\usepackage{amsthm}

\begin{document}

\title{Multi-bit quantum random number generation from a single qubit quantum walk}
\author{ Anupam Sarkar}
\affiliation{The Institute of Mathematical Sciences, C. I. T. Campus, Taramani, Chennai 600113, India}
\affiliation{Homi Bhabha National Institute, Training School Complex, Anushakti Nagar, Mumbai 400094, India}
\author{C. M. Chandrashekar}
\email{chandru@imsc.res.in}
\affiliation{The Institute of Mathematical Sciences, C. I. T. Campus, Taramani, Chennai 600113, India}
\affiliation{Homi Bhabha National Institute, Training School Complex, Anushakti Nagar, Mumbai 400094, India}


\begin{abstract}
We present a scheme for multi-bit quantum random number generation using a single qubit discrete-time quantum walk in one-dimensional space. Irrespective of the initial state of the qubit, quantum interference and entanglement of particle with the position space in the walk dynamics certifies high randomness in the system. Quantum walk in a position space of dimension $2^l+1$  ensures  string of $(l+ 2)$-bits of random numbers from a single measurement.  Bit commitment with the position space and control over the spread of the probability distribution in position space enable us with options to extract multi-bit random numbers. This highlights the {\it power of one qubit} , its practical importance in generating multi-bit string in single measurement and the role it can play in quantum communication and cryptographic protocols.  This can be  further extended with quantum walks in higher dimensions.
\end{abstract}

\maketitle

\section{Introduction}

Random number plays an important role in many applications where unpredictability is a key ~\cite{rubinstein2016simulation,metropolis1949monte}, especially in cryptographic protocols\,\cite{gisin2002quantum,ekert1991quantum,bennett1992experimental} where security is assured because of unpredictability. 
Though there are some statistical tests\,\cite{rukhin2001statistical,maurer1992universal,soto1999statistical} which can convice us about the random nature of the observed sequence, it is almost impossible to discriminate between a predetermined random string of bits that comes from a dishonest provider or malicious random number generator (RNG) and a {\it true} random sequence. In the first case the sequence may pass all the statistical tests but still can be completely predictable to the provider or anyone else who wants to eavesdrop. Therefore, generation of genuine randomness and its certification is generally considered impossible with only classical methods. Quantum physics brings out high unpredictability and probabilistic behaviour as an inherent property of nature\,\cite{born1955statistical}. Therefore,  one can expect certification of {\it true randomness} in quantum systems to come purely from the principles of quantum physics. 

The random nature of quantum mechanics\,\cite{bera2017randomness} has gained a lot of interest from the time of it's inception. Though the description of quantum system is probabilistic, the probabilistic prediction of a theory does not necessarily imply that it is intrinsically random. There can be some limitation to the formalism and a more complete theory can describe it in a completely deterministic way\,\cite{einstein1935can,bell2004speakable}. However, previous works\,\cite{acin2016certified,masanes2006general,barrett2005no} suggests that using the nonlocal correlation between two particles can generate the randomness which is truly intrinsic. For example, like measuring entangled particles one can assess the randomness of the process independent of it's quantum description which cannot be described deterministically within the framework of any no-signalling theories. Nonlocality has been proved as an important resource in many information processing tasks like random number generation protocols\,\cite{pironio2010random,ma2016quantum}, randomness expansion\,\cite{pironio2010random,colbeck2011private,coudron2014infinite} and amplification\,\cite{colbeck2012free,colbeck2009quantum} protocols, and quantum key distribution\,\cite{barrett2005no,acin2006bell,pironio2009device}. Though there is no direct  connection between nonlocality and entanglement\,\cite{brunner2014bell,horodecki2009quantum},  it is known that any pure entangled states are nonlocal. Using this nonlocality of observed statistics in bipartite Bell scenario, a device independent Quantum Random Number Generator (QRNG) has been suggested\,\cite{pironio2010random}. Other than that, various other approaches to built an efficient QRNG have been developed and all of them can be classified under three categories,  trusted device, self-testing, and semi self-testing \cite{ma2016quantum}. Though the device-independent or self-testing QRNG is more secure compared to two other protocols, it is unsuitable in some cases because of the slow generation of random numbers with time constrained under current technologies.

In this report we propose a QRNG solely based on the superposition and entanglement property of the quantum walk and we use pure states which associates the nonlocal behaviour as described before.  The motivation for using quantum walk is propelled by multiple advantages it can offer along with ability to generate multi-bit from a single qubit. 
The practical limit is bounded by the experimentally implementable number of steps of quantum walks in any system like, NMR\,\cite{ryan2005experimental},trapped ions\,\cite{schmitz2009quantum,zahringer2010realization}, cold atoms\,\cite{karski2009quantum}, and photonic systems\,\cite{peruzzo2010quantum,schreiber2010photons,perets2008realization,broome2010discrete}. 
Our analytical and numerical analysis shows that the randomness of an initial state of the particle is being enhanced using the quantum walk dynamics. The result suggests that it's dependency on the initial state is very weak and this ensures that a significantly high randomness is seen even when randomness in initial state is zero.


\noindent
{\bf{ Discrete-time quantum walk}} : The Discrete Time Quantum Walk (DTQW) is defined on the Hilbert space $\mathcal{H}= \mathcal{H}_{c}\otimes \mathcal{H}_{p}$ where 
$\mathcal{H}_{c}$ is the Hilbert space of the particle/walker and $\mathcal{H}_{p}$ is the position Hilbert space\,\cite{riazanov1958feynman, feynman1986quantum, meyer1996quantum, aharonov1993quantum, venegas2012quantum, chandrashekar2011disordered, chandrashekar2012disorder,singh2017interference}. In this paper we consider one-dimensional DTQW with the  particle having two internal degrees of freedom. Therefore, $\mathcal{H}_{c}$ is spanned by the basis states $\lbrace\ket{\uparrow}, \ket{\downarrow}\rbrace$ and we will call it coin space. For the position space the basis states will be $\lbrace\ket{i} : i\in  \mathbb{Z} \rbrace$. Each step of DTQW comprises of quantum coin operation,
\begin{align}
C(\theta)= \begin{bmatrix}
   ~~\cos\theta & -i\sin\theta \\
   -i\sin\theta & ~~\cos\theta
 \end{bmatrix}
\end{align}
followed by a position shift operator defined as 
\begin{align}
\mathcal{S}_{x}\equiv\sum_{x} \Big[\ket{\uparrow}\bra{\uparrow}\otimes \ket{x-1}\bra{x}+\ket{\downarrow}\bra{\downarrow}\otimes \ket{x+1}\bra{x}\Big].
\end{align}
The resulting operation $\mathcal{W}(\theta)= [\mathcal{S}( C(\theta)\otimes\mathds{1})]$ evolves the particle in superposition of position space which has no classical analogue and quite advantageous for many information processing tasks and an integral part of quantum simulation schemes. The state of the walker after $t$-steps will be $\ket{\psi_{t}}= \mathcal{W}(\theta)^{t}\ket{\psi_{in}}$, where $\ket{\psi_{in}}$ is the initial state of the walker or the particle. In our consideration \begin{equation}\label{initial}
\ket{\psi_{in}}= \Big(\cos{\delta}\ket{\uparrow}+e^{i\eta}\sin{\delta}\ket{\downarrow}\Big)\otimes \ket{x=0}.
\end{equation}
Using this initial state we will study the behaviour of the randomness under quantum walk dynamics.

\section{Results}

{\bf{Randomness in coin and position space}} : Here, we consider a specific form of quantification of randomness in a quantum system termed as intrinsic randomness of measurement\,\cite{yuan2015intrinsic}. It has been quantified as a coherence measure and clarifies the operational aspect of quantum coherence. Since the QRNG protocol we propose is solely based on DTQW dynamics, it is necessary to have a good measure of randomness contained in both position and coin space. For evaluating the randomness associated with coin space we have to trace out the part of Hilbert space associated with position space from the density matrix and from the reduced density matrix we can compute the randomness as described in Methods section. Similarly, by tracing out the coin space we can calculate the randomness incorporated with the position space. One important point to note here is that the randomness computed is explicitly of quantum origin and is of different nature from any randomness originated through classical stochastic process.\\ 
\\
\noindent
{\it{Randomness in the initial state~}} : If the initial state of walker is of the form given in Eq.\,\eqref{initial}
 (we will omit $x$ whenever no confusion arises), corresponding density matrix would be 
 \begin{align}
 \rho_{in}=& \ket{\psi_{in}}\bra{\psi_{in}} \nonumber
 =(\cos^{2}{\delta}\ket{\uparrow}\bra{\uparrow}+e^{-i\eta}\sin{\delta}\cos{\delta}\ket{\uparrow}\bra{\downarrow} \nonumber \\
 &+e^{i\eta}\sin{\delta}\cos{\delta}\ket{\downarrow}\bra{\uparrow}+ \sin^{2}{\delta}\ket{\downarrow}\bra{\downarrow})\otimes \ket{0}\bra{0}.
\end{align} 
Using the preceding expression it is easy to calculate the randomness associated with coin and position space individually. For coin space it would be dependent on the parameter $\delta$ only and can be expressed as,
\begin{align}
  R_{i}(\rho_{in}^{c})= -(\cos^{2}(\delta)\ln(\cos^{2}(\delta))+\sin^{2}(\delta)\ln(\sin^{2}(\delta)).
  \end{align}
Since the walker initially is fixed at one position their is no inherent randomness associated with position. Thus, randomness associated with initial position space will be $0$. This  supports  the viability of randomness quantification process.\\

{\bf{Randomness after $t$- steps~}} : After $t$-steps, the generic form of the walker can be written as $\ket{\psi_{t}}= (\sum_{x}a_{x,t}\ket{\uparrow}+b_{x,t}\ket{\downarrow})\otimes\ket{x}$, which is the outcome of the operation $\mathcal{W}(\theta)^{t}$ on the initial state. Therefore, the density matrix corresponding to the state would be,
\begin{align}
\rho_{t} & = \ket{\psi_{t}}\bra{\psi_{t}}
= \sum_{x,y}(a_{x.t}a_{y,t}^{*}\ket{\uparrow}\bra{\uparrow}+a_{x,t}b^{*}_{y,t}\ket{\uparrow}\bra{\downarrow} \nonumber \\
& +b_{x,t}a^{*}_{y,t}\ket{\downarrow} \bra{\uparrow}  +b_{x,t}b^{*}_{y,t}\ket{\downarrow}\bra{\downarrow})\otimes\ket{x}\bra{y}.
\label{rhodef}
\end{align}
Now using this expression we can compute the randomness associated with the position and coin space individually and both together.\\

\begin{figure}
 \includegraphics[width=9cm]{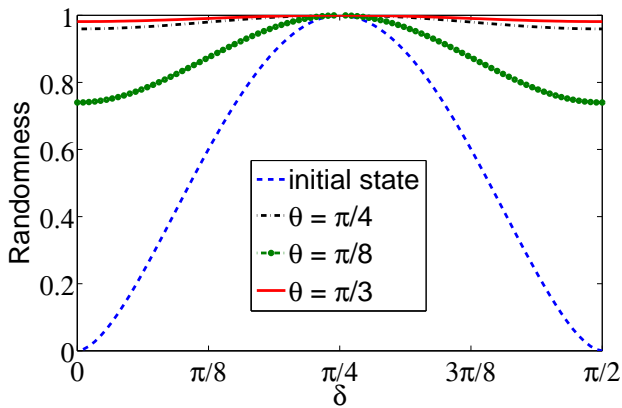} 
    \caption{Intrinsic randomness in the coin space (particle) as a function of initial state parameter $\delta$ before implementing quantum walk and after implementing 50 step of quantum walk using difference coin operation parameter $\theta$.  Though we see some dependency on $\delta$, a significant enhancement of randomness is seen even when the randomness in the initial state is zero.}
\label{fig1}
\end{figure}

{\it{Randomness in coin space}} : The walker has two internal degrees of freedom and total randomness is being distributed in the form of probability amplitude associated with the $\ket{\uparrow}$ and $\ket{\downarrow}$ states. By tracing out the position space, we will be remained with the reduced density matrix denoted by $\rho_{in}^{c}$ expressed in the form, $\rho_{t}^{c}= \rho_{11}\ket{\uparrow}\bra{\uparrow}+\rho_{12}\ket{\uparrow}\bra{\downarrow}+\rho_{21}\ket{\downarrow}\bra{\uparrow}+\rho_{22}\ket{\downarrow}\bra{\downarrow}$. So, the randomness can be expressed as
 \begin{align}
 R_{i}(\rho_{t}^{c}) & = -(\rho_{11}\ln\rho_{11}+\rho_{22}\ln\rho_{22}) \nonumber \\
 &= \sum_{x}\vert a_{x,t} \vert ^{2}\ln \vert a_{x,t}\vert^{2}+ \vert b_{x,t}\vert^{2}\ln\vert b_{x,t}\vert^{2}.
\end{align} 
In Fig.\,\ref{fig1} we show the randomness in the coin space as function of $\delta$ which fixes the initial state before the walk and after 50 step of walk using different coin operation parameter $\theta$.  Irrespective of the initial state, that is, even when initial state's randomness is zero,  we can see an high value of randomness after 50 steps of DTQW. From the analytical results presented in the "Supplementary Information" and the numerical results we can say that the same behaviour will be seen even after small number of steps.\\

\begin{figure}
\includegraphics[width=9cm]{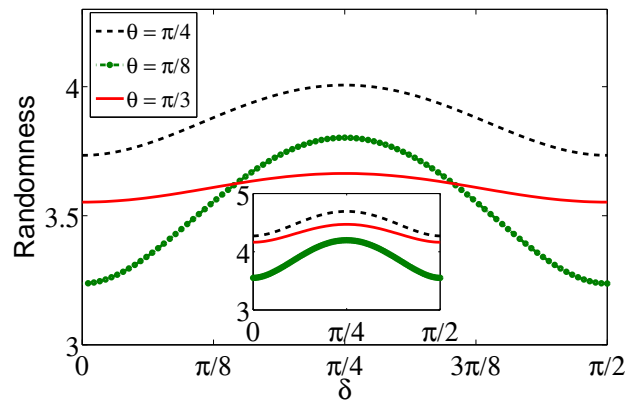}
\caption{Intrinsic randomness in position space as a function of initial state parameter $\delta$ after 25 step of walk using different coin operation parameter $\theta$. It is evident from the figure that randomness shows some dependency on the initial state, however,  the variation is very small when compared to the zero randomness in the initial stage. Inset in the figure is the randomness when both coin and position space are taken together. We see a unit increase in the randomness when both the space are taken into account.}
\label{fig2}
\end{figure}

{\it{Randomness in position space}} : We can use the randomness of the state extended in superposition of  position space to extract intrinsically random classical bit string out of it. Extraction process is discussed in {\it{extracting randomness}} section following this. For quantification of randomness in position space we will follow the same recipe as we used in quantifying randomness in coin space. If the dynamics of the walker involves $t$ number of steps of walk then we know that generic state can be written as $\ket{\psi_{t}}= \sum_{x}(a_{x,t}\ket{\uparrow}+b_{x,t}\ket{\downarrow})\otimes\ket{x}$, and $\rho_{t}= \ket{\psi_{t}}\bra{\psi_{t}}$. By tracing out the coin space we will get the form as $\rho_{t}^{p}= \sum_{i, i^{'}}\rho_{i,i^{'}}\ket{i}\bra{i^{'}}$. To calculate the randomness inherited by the reduced state, we need the diagonal entries in computational basis that is,  $\rho_{i,i}$.  The density matrix after time $t$ can be written in the form, 
\begin{align}
\rho_{t} &= \ket{\psi_{t}}\bra{\psi_{t}} = \sum_{x,y}(a_{x.t}a_{y,t}^{*}\ket{\uparrow}\bra{\uparrow}+a_{x,t}b^{*}_{y,t}\ket{\uparrow}\bra{\downarrow} \nonumber \\
& +b_{x,t}a^{*}_{y,t}\ket{\downarrow
}\bra{\uparrow}+b_{x,t}b^{*}_{y,t}\ket{\downarrow}\bra{\downarrow})\otimes\ket{x}\bra{y}
\end{align}
therefore 
\begin{align}
\rho_{t}^{p} = \sum_{x,y=-t}^{t}(a_{x,t}a_{y,t}^{*}+ b_{x,t}b_{y,t}^{*})\otimes \ket{x}\bra{y}.
\end{align}
 Comparing the two expression for $\rho_{t}^{p}$ we can write down the form of the randomness as 
\begin{align}
R_{i}({\rho_{t}^{p}})= \sum_{x}(\vert a_{x,t}\vert^{2}+\vert b_{x,t}\vert^{2})\ln(\vert a_{x,t}\vert^{2}+\vert b_{x,t}\vert^{2}).
\end{align}
In Fig.\,\ref{fig2}, randomness in position space after  25 step of quantum walk as a function of initial state parameter $\delta$ is shown. Before the walk, randomness in position space is zero therefore, what we note after 25 steps of walk is a significant quantity of randomness in the system even though it varies a bit as function of $\delta$ and for different value of $\theta$.  In the inset of Fig.\,\ref{fig2}, randomness in complete system, coin and position space together is shown. We see an overall  boost in the randomness when compared to randomness in position space alone. 
\\ 



{\bf{Extracting randomness~}}: The working principle of a QRNG is to make use of the quantum phenomena such as superposition of quantum states and measurement to obtain classical output string which is desirable to be random enough to pass any statistical tests. It is known before measuring that, a two-level system can acquire random classical bit string from the random outcome of the measurement. We will see how a qubit quantum walk can generate single random bit and a multi-random bit string and its advantages over using a single copy of qubit system.

{\it{ From coin space~}} : We have already discussed the procedure to quantify randomness in coin space and our analysis shows it's dependence on the initial state of the walker. With evolution of the walk, a clear enhancement of the randomness in comparison to the initial state is seen. That is, after $t$- steps the particle evolves according to the dynamics and probability amplitude corresponding $\ket{\uparrow}$ and $\ket{\downarrow}$ states will keep changing  after each step. To get the random classical bit out of it we have to use a detector to detect the state of the walker after any arbitrary number of steps. Here we will trust both, the device implementing quantum walk and detector to get intrinsically random classical bit. Walker will be in the superposition state $a_{x,t}\ket{\uparrow}+b_{x,t}\ket{\downarrow}$ before the detection and it would  collapse on either of these two states after measurement and we will code a classical bit $0$ or $1$ for detecting $\ket{\uparrow}$ and $\ket{\downarrow}$ state, respectively. Here, the bit commitment is arbitrary and we could use the opposite commitment too. Since quantum mechanics assures us the outcome being inherently random, we cannot  have a prior knowledge about the outcome before detection. Therefore, we can expect a perfectly random classical series of string as output by repeating this scheme for several times.

{\it{ From position space~}} : We will use the same kind of extraction process in the position space as in the coin space. The advantage of using the position space is its ability to generate a multiple-random bit string rather than a single bit after each round of extraction like it is in coin space. More explicitly, after $t$-steps walk, tracing out the coin space, we can write the generic state in position space in the form $\ket{\psi_{t}}= \sum_{x=-t}^{t}a_{x,t}\ket{x}$, which is in superposition of all possible states corresponding to each position. To make a measurement we have to place a position resolving detectors (or a multiple - detector at each position) where the particle will be detected. If we use the standard version of DTQW then, 
\begin{align}
 \ \ket{\psi_{t}} &=  a_{t,t}\ket{t}+ a_{t-2,t}\ket{t-2}+ a_{t-4,t}\ket{t-4}+ \cdots \nonumber  \\  
 & +a_{-t+2, t}\ket{-t+2}+a_{-t,t}\ket{-t}.
\end{align}
We can define the state of the detector using a simple mathematical formula, $2^{n}= 2t$. If $n$ is an integer then we will use $n$ number of quantum bits to denote the state of the detectors and if $n$ is not an integer then the maximum number of quantum bits needed to specify all detectors is $\lbrace n\vert \min\limits_{n}2^{n}\geq 2t\rbrace$. If $2t=2^{n}$ then the detector states will be defined as follows :
\begin{center}
\begin{tabular}{ |p{3cm}||p{5cm}|  }
 \hline
 \multicolumn{2}{|c|}{Bit commitment scheme with position} \\
 \hline
 Detected position of walker & Corresponding state\\
 \hline
 ~~~~~~~$-t$  & $\ket{00\cdots0}$, $0$ appears $n$ times   \\
 ~~~~~~$-t+1$ &  $\ket{00\cdots1}$,$0$ appears $n-1$ times  \\
 ~~~~~~~~~~\vdots & ~~~~~~~~~\vdots \\
 ~~~~~~~~~~\vdots & ~~~~~~~~~\vdots \\
 ~~~~~~ $t-1$ & $\ket{11\cdots0}$, $1$ appears $n-1$ times\\
 ~~~~~~`~~~$t$ & $\ket{11\cdots1}$, $1$ appears $n$ times. \\
 \hline
\end{tabular}
\end{center}
Therefore, if the particle is being detected at position $t$ then we will note the $n$ bit string associated with the detector at which it is measured. Below we present a table with an example of bit commitment scheme after  $8$ step of quantum walk 
\begin{center}
\begin{tabular}{ |p{3cm}||p{3cm}|  }
 \hline
 \multicolumn{2}{|c|}{Bit commitment scheme after 8 step of walk} \\
 \hline
 Detected position of walker & State of the walker\\
 \hline
 ~~~~~~~$-8$  & ~~~~~~~~~$\ket{0000}$    \\
 ~~~~~~~$-7$ &  ~~~~~~~~~$\ket{0001}$ \\
 ~~~~~~~$-6$ & ~~~~~~~~ $\ket{0010}$\\
 ~~~~~~ $-5$ & ~~~~~~~~~$\ket{0011}$\\
 ~~~~~~~$-4$ & ~~~~~~~~~$\ket{0100}$ \\
 ~~~~~~~$-3$ & ~~~~~~~~~$\ket{0101}$\\
 ~~~~~~~$-2$ & ~~~~~~~~~$\ket{0110}$\\
 ~~~~~~~$-1$ & ~~~~~~~~~$\ket{0111}$\\
 ~~~~~~~~~$1$  & ~~~~~~~~~$\ket{1000}$    \\
 ~~~~~~~~~$2$ &  ~~~~~~~~~$\ket{1001}$ \\
 ~~~~~~~~~$3$ & ~~~~~~~~ $\ket{1010}$\\
 ~~~~~~~~~$4$ & ~~~~~~~~~$\ket{1011}$\\
 ~~~~~~~~~$5$ & ~~~~~~~~~$\ket{1100}$ \\
 ~~~~~~~~~$6$ & ~~~~~~~~~$\ket{1101}$\\
 ~~~~~~~~~$7$ & ~~~~~~~~~$\ket{1110}$\\
 ~~~~~~~~~$8$ &~~~~~~~~~$\ket{1111}$\\
 \hline
\end{tabular}
\end{center} 
Here we have not  committed any  bit string for position $0$. However, we can assign a bit $0$ or $1$ arbitrarily or we can avoid committing a bit string for this or any specific position according to the choice of the client who wants to generate the random number. In DTQW evolution we know that after odd (even) number of steps of walk, positions identified with even (odd) number will have zero probability of finding a particle. This will eliminate the occurrence of half of the configuration of multi-bit random number. To address this concern we can use split-step quantum walk \,\cite{kitagawa2010exploring,mallick2016dirac} or directed quantum walk\,\cite{hoyer2009faster,chandrashekar2014quantum}.

\begin{figure}
\includegraphics[width=9cm]{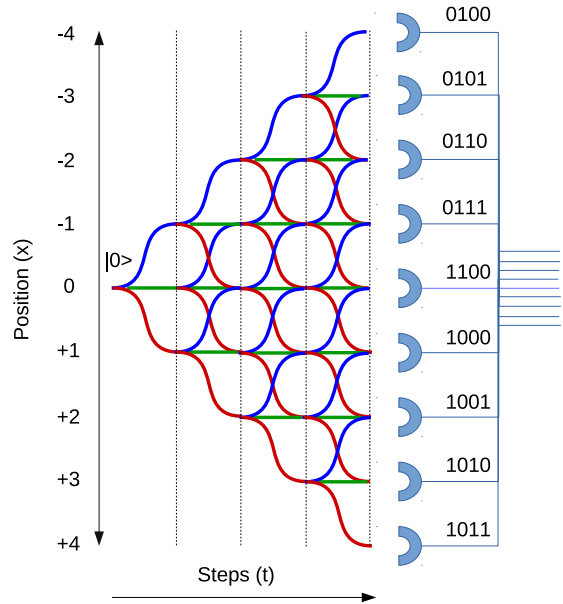}
\caption{Schematic representation four bit random number generation from a single qubit after four step of SS-QW. If the state of the walker is also considered, we will effectively have five bit random number after measurement.}
\label{fig3aa}
\end{figure}

\noindent
{\bf Split-step quantum walk} : In a one-dimensional split-step quantum walk (SS-QW) the shift operator is divided into two parts denoted by $S_{-}$ and $S_{+}$. These operations are defined as
\begin{align}
 S_{-}= \sum_{x}\ket{x-1}\bra{x}\otimes\ket{\uparrow}\bra{\uparrow}+\mathds{1}\otimes\ket{\downarrow}\bra{\downarrow} ; \nonumber \\
S_{+}= \sum_{x}\mathds{1}\otimes\ket{\uparrow}\bra{\uparrow}+\ket{x+1}\bra{x}\otimes\ket{\downarrow}\bra{\downarrow}.
\end{align} 
 Unlike standard form of DTQW,  here two different coin operators dependent on two different parameters 
 $\theta_{1}$ and $\theta_{2}$ are used. Therefore, the resulting operation will be defined as,
\begin{align}
\mathcal{W}(\theta_{1}, \theta_{2})= [\mathds{1}\otimes C(\theta_{2})]S_{-}[\mathds{1}\otimes C(\theta_{1})]S_{+},
\end{align}
where for the coin operator we used the same form as above. After $t$ steps of walk the state will be 
\begin{align}
\ket{\psi_{t}}= \mathcal{W}(\theta_{1}, \theta_{2})^{t}\ket{\psi_{in}}
\end{align}
and it can be written in the form $\ket{\psi_{t}}= \sum_{x=-t}^{t}(a_{x,t}\ket{\uparrow}+b_{x,t}\ket{\downarrow})\otimes\ket{x}$. Here the probability amplitude will be non-zero for each position and state will be of the form, 
 \begin{align}
 \ket{\psi_{t}} &= a_{t,t}\ket{t}+ a_{t-1,t}\ket{t-1}+a_{t-2,t}\ket{t-2}+\cdots\nonumber \\ & +a_{-t+1,t}\ket{-t+1}+a_{-t,t}\ket{-t}.
\end{align}
We can calculate the randomness using the pure state density matrix and by tracing out the coin space as we did in standard DTQW case. In Fig.\,\ref{fig3aa} we show the schematic representation of generation of four bit random number after  four step of SS-QW.\\
\noindent
 {\bf Directed quantum walk}: To define the directed discrete time quantum walk (D-QW) in one dimension, we will be using one directed edge connecting two vertices of the graph and $(n-1)$ self looping edges at each vertex and we will assign a basis vector to each edge. Then every state at each edge can be expressed as linear combination of the states $\ket{x, \rightarrow}, \ket{x,1},\ket{x,2}\cdots\ket{x,n-1}$ where $x$ is non-negative integer and $\rightarrow$ indicates edge along the line and each number comes from distinct self loop. The action of shift operation is defined as $S_{x}\ket{x,\rightarrow}= \ket{x+1,\rightarrow}$ and $S_{x}\ket{x,i}= \ket{x,i}, i\in [1,n-1]$. Therefore, the shift operator takes the  form 
 \begin{align}
 S_{x}= \sum_{x,i}\ket{\rightarrow}\bra{\rightarrow}\otimes\ket{x+1}\bra{x}+\ket{i}\bra{i}\otimes\ket{x}\bra{x}.
 \end{align}
Coin operation has the form  $C\equiv\begin{bmatrix}
 \alpha & \beta\\
 \beta & -\alpha                                                                                                                                                                                                                                                                                                                                                                                                                                                                                                                                                                                                                                                                                                                                                                                                                                                                                                                                                                                                                                                                                                                                                                                                                                                                                                                                                                                                                                                                                                                                                                                                                                                                                                                     \end{bmatrix}$ where $\alpha = 1/\sqrt{n}$ and $\beta = \sqrt{\frac{n-1}{n}}$. Therefore, one step of D-QW comprises of two operation $\mathcal{W}_{d}= S_{x}[C\otimes\mathds{1}]$. After $t$ steps of D-QW the state will be $\ket{\psi_{t}}= \mathcal{W}_{d}^{t}\ket{\psi_{in}}$.
  \begin{align}
 \ \ket{\psi_{t}} &=  a_{0,t}\ket{0}+ a_{1,t}\ket{1}+ a_{2,t}\ket{2}+ \cdots \nonumber
 \\& +a_{t-1, t}\ket{t-1}+a_{t,t}\ket{t}.
\end{align}

\begin{figure}
\includegraphics[width=9cm]{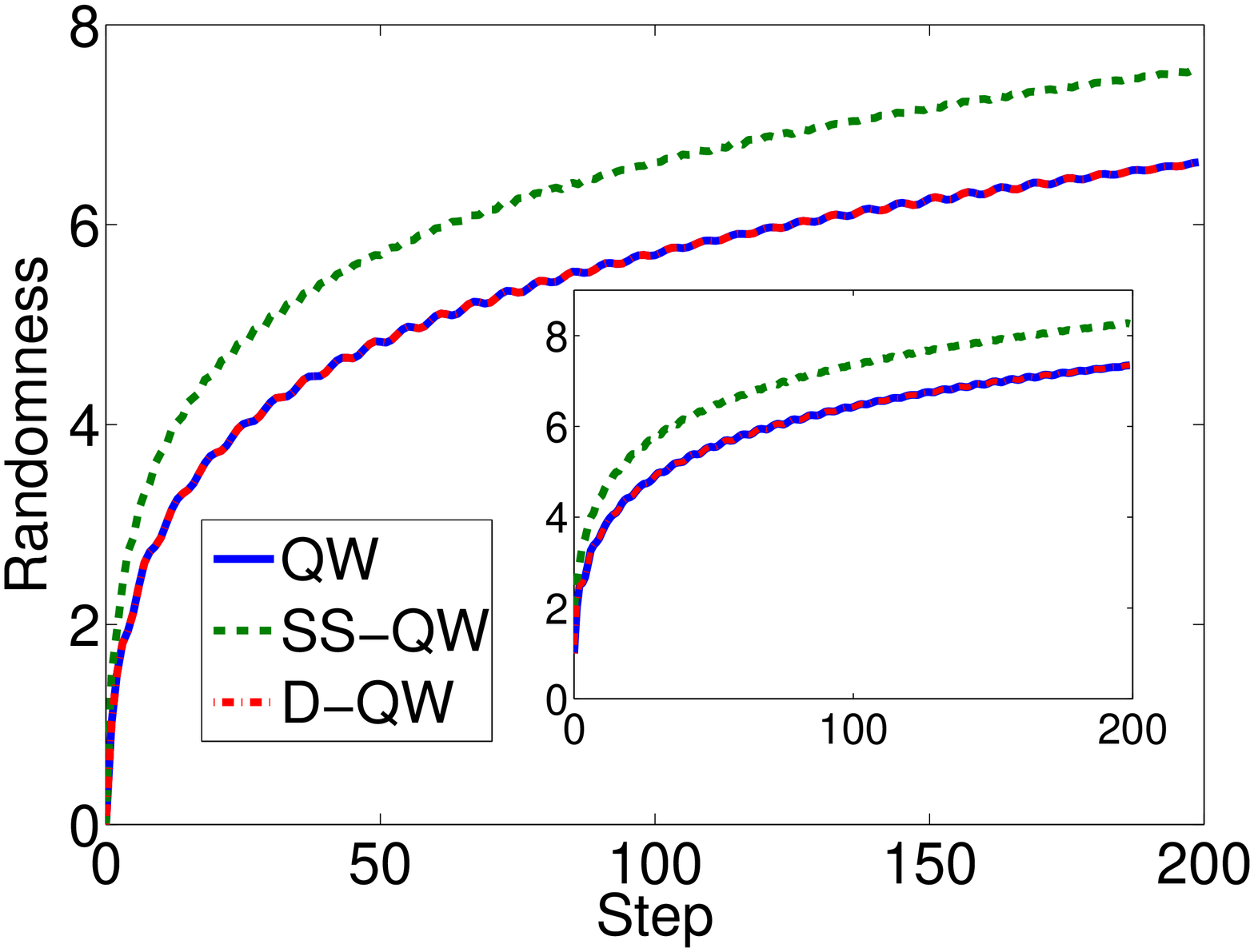}
\caption{Randomness with number of steps in standard DTQW, SS-QW, and D-QW.  Due to non-zero probability at all position space in SS-QW we can see a maximum randomness compared to other two. Amount of randomness measure corresponds to equivalent number of qubit it can mimic in the process. Inset in the figure is the randomness when both coin and position space are taken together.}
\label{fig3}
\end{figure}

In Fig.\,\ref{fig3}, randomness in the system as function of number of steps is shown when measurement are made only in the position space and in both, position and coin space (inset).  {\it The increase in randomness shows the way number of equivalent quantum bit the system mimics using a single qubit.}  The plot shows the randomness for all three types of quantum walk evolution. Due to non-zero probability at all position space, the randomness is more (maximizes)  for the SS-QW than the randomness obtained for standard DTQW (Fig.\,\ref{fig3}). In D-QW the number of positions on which probability amplitude is non-zero is equivalent to number of position space with non-zero probability in standard DTQW, the randomness measure is also identical (Fig.\,\ref{fig3}). In the inset we have shown the randomness in both, coin and position space together. Inclusion of coin space enhances the randomness in the system by a small amount.\\

{\it Advantages of using both space} : If we take into account both, the coin and position degrees of freedom,  we can generate an extra bit compared to the string of bits from position space alone. The reason is very obvious as we have seen already how internal degrees of freedom of the particle is able to generate one classical random bit after a single round of evolution. Therefore, when we use both, we will get random classical bit string coming from both spaces. The extraction process is same as discussed before but the detectors  at each possible positions should be capable of capturing the information about internal state of the walker along with detecting the position of the walker.

In this scheme using dynamic evolution of qubit,  from a given position space we can generate a uniform length classical multi-bit string rather than a single bit from a standard single particle QRNG. The randomness depends only on the device's trustedness. 

{\it{Randomness quantification under noise}}: An important part of any cryptographic protocol is to ensure it's security under many possible attacks. It is in general impossible to make it secure under any arbitrary attacks but we can prove its security considering some realistic cases. One of these cases will be the difficulty to create a perfect pure state as different kind of noises will mix it up resulting a mixed state. Now the concern is that an Eavesdropper can have access of these noises and he or she might be able to get some information about the measurement outcomes using this correlation with the system. This is highly undesirable as we want to use this measurement outcome as random number in other cryptographic protocols. Let us consider the case where instead of starting with pure state some noise is mixed and we will consider a purified state $\ket{\psi^{CPE}}$ which is being shared by the walker and the adversary Eve. This is the state in the larger Hilbert space $\mathcal{H_{C}}\otimes\mathcal{H_{P}}\otimes\mathcal{H_{E}}$ where the walker does not have the access to the Hilbert space $\mathcal{H_{E}}$. Purification implies that the state of the walker will be  $\rho^{CP}= Tr_{E}(\ket{\psi^{CPE}}\bra{\psi^{CPE}})$. Upon obtaining the outcome $e$ with probability $p_{e}$ by doing projective measurement on Eve's system, the state of the walker will be $\ket{\psi_{e}^{CP}}= \langle \psi_{e}^{E}|\psi^{CPE}\rangle$. The measurement of randomness corresponding to the walker's state would be $R_{i}(\ket{\psi_{e}^{CP}})$ and the total randomness can be quantified as $\sum_{e}p_{e}R_{i}(\ket{\psi_{e}^{CP}})$. Now Eve's optimal strategy would be to choose a measurement basis which will maximize her side information about the measurement outcome, equivalently saying minimizing the randomness of the walker's measurement outcome. Therefore, the total randomness can be quantified as 
\begin{align}
R_{i}(\rho^{CP})= \sum_{min (p_{e},\psi_{e}^{E})}p_{e}R_{i}(\ket{\psi_{e}^{CP}}).
\end{align}

\begin{figure}
\includegraphics[width=9cm]{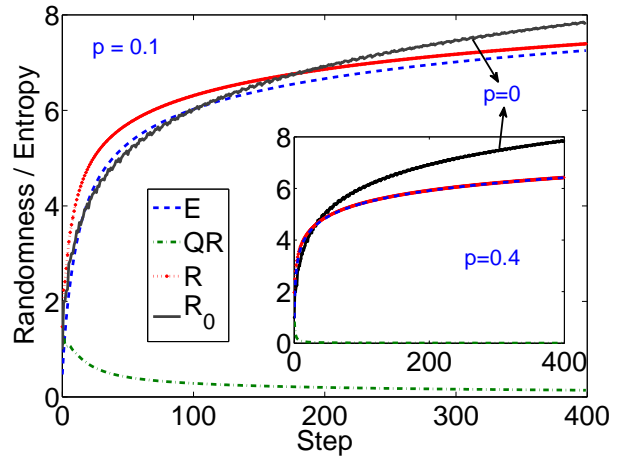}
\caption{Randomness with number of steps in standard DTQW in presence of bit flip noise in coin space. With increase in noise some decrease in randomness is seen. The randomness in system with noise will have contribution from both, quantum origin as well as noise. To show that we have plotted randomness ($R$) for noiseless evolution and for evolution with different noise level $p$. The van Neumann entropy ($E$) and the randomness of quantum origin is also shown. Though overall randomness does not see a significant decrease, a substantial decrease in randomness from quantum origin  is seen.}
\label{fig3aB}
\end{figure}
An other description of effect noise on the randomness in the system will be in the form of decoherence in DTQW evolution.  A simple form of introducing decoherence into DTQW evolution will be in the form of bit flip noise, that is,  $\rho(t)$ of the complete system as outlined in Eq.\,\eqref{rhodef} after every time step $t$ will be in the form  $\rho(t) = p[(\sigma_x\otimes \mathds{1}) \rho(t)(\sigma_x\otimes \mathds{1})^{\dagger}] + (1-p)\rho(t)$. Here $p$ is the noise level and $\sigma_x$ is bit flip operation.  This bit flip noise results in decrease in spread of the wavepacket in position and correspondingly the randomness in the system also decreases. However, when we calculate the randomness after such evolution we will have contribution from both, quantumness in the system and from the noise process. Therefore, while using DTQW with noise, all the randomness we obtain cannot be attributed to the quantum origin. To give a quantitative picture of the randomness of quantum origin in DTQW system, in Fig.\,\ref{fig3aB} we present a plot of extractable randomness in complete DTQW system with noise of different level and the randomness of quantum origin for the corresponding noise level. 

When their is no noise in the dynamics ($p=0$, $R_0$) we see a study increase in randomness with number in steps and all the randomness can be attributed to the quantum origin. With increase in noise level ($p=0.1$ and $p=0.4$, $R$) we first seen a steep increase in randomness and with time it will be lower than the randomness of quantum origin when $p=0$. For non-zero $p$ we have show the value of von Neumann entropy, $E(t) = -tr [\rho(t) \ln (\rho(t)]$, a good measure of randomness due to noise in the system.  Therefore, the contribution of randomness of quantum origin can be quantified in the form $QR(t) = R(t) - E(t)$.  This value  becomes very small with increase in $p$ and time $t$ and for $p=0.4$ its almost zero.  We should note that this method of evaluating randomness of quantum origin is valid only for the system with $\rho(t)$ is a pure state when their is no noise (that is, when $E(t) = 0$ for $p=0$). However, though contribution of randomness from quantum origin decrease with in increase in noise level, the overall randomness of the system will continue to scale the same way as the spread of the DTQW scales with noise.   

{\it Randomness as guessing probability} : Out of many possible attacks one can consider is the fact that the detectors are correlated with an adversary's system.  Upon obtaining a specific outcome, the state of the adversary would be changed accordingly and measuring the correlated system in her lab, Eve can get some useful information about the outcome of the QRNG. However,  considering our case, the detectors are placed at every possible positions where the particle can be detected and corresponding state can be written as $\ket{i}$ where $i$ is a $n-$bit string as we suggested in the extraction procedure.

If the probability distribution of the random variable (measurement outcome) being denoted by $P_{I}$ then consider an adversary whose states $\rho_{i}^{E}$ depends on the random variable $I$ which corresponds to the Classical-Quantum state $\rho_{PE}:= \sum_{i\in I}P_{I}(i)\ket{i}\bra{i}\otimes \rho_{E}^{i}$. If  $P(I|E)= \sum_{i\in I}P_{I}(i)Tr (\pi_{i}\rho_{i}^{E})(\pi_{i}$ is the POVM on Eve's system$)$, it is the probability of guessing the outcome of position measurement using the optimal measurement strategy by the adversary. For this  it has been proved\,\cite{konig2009operational} that $P(I|E)$ is related to the $min-$ entropy by the relation $P(I|E)= 2^{-H_{min}(I|E)}$ where the $min-$ entropy is defined by $H_{min}(I|E):= -\inf_{\sigma_{E}}D_{\infty}(\rho_{PE}||\mathds{1}_{P}\otimes\sigma_{E})$ and $ D_{\infty}(\rho|\sigma):= \inf\lbrace\lambda\in\mathbb{R}: \rho\leq 2^{\lambda}\sigma\rbrace$.

\section{Methods}

{\it{ Randomness quantification~}}: The intrinsic randomness of a quantum system is related to the random outcomes of the measurement on the system\,\cite{yuan2015intrinsic}. If we measure a pure state $\rho=\ket{\psi}\bra{\psi}$, (where $\vert\psi\rangle = \sum_{i}a_{i}\vert i \rangle$)
in the basis $\vert i \rangle$ considering projective measurement, then measurement outcomes are intrinsically random.
\\
According to Born's rule  $ p_{i}= Tr[P_{i}\rho]= \langle i \vert\rho\vert i \rangle = \langle i \vert \sum_{j,k} a_{j}a_{k}\vert j\rangle\langle k \vert \vert i \rangle = \vert a_{i}\vert^{2}$ will be the probability of obtaining the $i$'th outcome. $P_{i}$ are the rank one projectors on the basis states.
 Then randomness of the output random variable is defined as
$$ R_{i}(\rho=\vert\psi\rangle\langle\psi\vert) = -\sum_{i} p_{i} \ln p_{i}.$$
Which is the Shannon entropy function of the probability distribution $\lbrace p_{i}\rbrace$. In another way it can be written as $ R_{i}(\rho=\vert\psi\rangle\langle\psi\vert) = S(\rho^{diag})$, where $\rho^{diag}$ is the density matrix that has only diagonal terms of $\rho$ in the computational basis $\lbrace i \rbrace$. If we think $\rho$ as a $n\times n$ matrix, having only diagonal terms $\rho_{ii}$ in the computational basis then the randomness inherited by state $\rho$ can be quantified as 
\begin{align}
R_{i}(\rho)= -\sum_{i=1}^{n}\rho_{ii}\ln\rho_{ii}.
\end{align}
From the preceding expression we can note that the information of diagonal elements which corresponds to probability distribution of each of the two basis state quantum walk in position space is sufficient to obtain the randomness in the system. Therefore, an analytical expression for $a_{x, t}$ and $b_{x, t}$, amplitudes of the basis state $|\uparrow\rangle$ and $|\downarrow\rangle$ at position $x$ and time $t$ for noiseless and decoherent quantum walk can be obtained from the Fourier analysis of the walk as described in Ref.\,\cite{NV00} and \cite{BCA03}, respectively. Using the first and the second moments, analytical expression for the asymptotic behaviour has also been presented in the same references.

\section{Discussion and Conclusion}

Many of the existing  protocols for Quantum Random Number Generator (QRNG) are based on quantum state preparation and measurement schemes. We introduce here a QRNG based on quantum dynamics which can be controlled rather than the  "prepare and measure" methods.  In addition to that, our paper includes the following important results and advantages to develop a QRNG protocol using quantum walk.
\\
For Discrete Time Quantum Walk (DTQW), the quantification process for the randomness inherited by the state (for both, pure and mixed) has been prescribed.
\\ After implementing the DTQW, we have analysed the dependency of  the quantified intrinsic randomness on the initial state parameter $\delta$ and walk evolution parameter $\theta$. We have provided the analytical calculation (see Appendix) for short time and numerical results for long time evolution to prove the fact that in both, position and coin space, the randomness is almost independent of the walk evolution parameter $\theta$.  Though the randomness in position space shows some dependence on the initial state parameter $\delta$, the degree of dependence is very small compared to the initial state randomness, which is being calculated as zero.  A significant enhancement of randomness with the increase in number of steps of DTQW is an other important thing that has been highlighted. 
\\ We have suggested a scheme to extract random number both, from measurement on coin space and position space  outcomes individually and together. Our extraction scheme shows the use of single particle and incorporating position degrees of freedom with it to generate a bit string of random number after measurement. We have also  established that the long bit string out of single run of the single particle system  can be obtained by increasing the dimension of position Hilbert space. This clearly implies that the bit-rate can be made higher depending on the position space dimension we can control under current technologies. Many earlier results have demonstrated ways to control probability distribution of the DTQW\,\cite{TFM03, CSL08, ALB17, kumar2018bounds, PF18, singh2018accelerated, Gio19}. Therefore, the probability distribution of the walk can be engineered to get a uniform probability distribution which is desired in any cryptographic protocols or any other distribution. This helps us to design a QRNG protocol which is capable enough to give us the random numbers from a desired probability distribution and any undesired nature will directly indicate the presence of adversary or hardware failure.

The main idea of our work was to construct a protocol for generating multiple random bits using quantum walk of a single particle. Recent experimental demonstration of using position degree of freedom along with polarization and orbital angular momentum degree of freedom, a muti-qubit entangled state has been reported\,\cite{Wan18}. Therefore, realization of multi qubit state using extended position space is not far from experimental feasibility.  We have also shown the advantages of using SS-QW over D-QW or standard DTQW in extracting higher randomness with all possible combination of multi-bit string. This behaviour is expected because for the SS-QW where the degrees of freedom in position space with non-zero probability is double.  In comparison with other two,  the probability amplitude at  all the possible positions using SS-QW effectively contribute to the expression of randomness. 

Now the question can arise about the probability distribution of the position space, where uniformity of the distribution is desired for the security purpose. Controlled distribution of quantum walk can in general be engineered to pick the desired distribution using position dependent coin operations and along with phase operations. Recently, an experimental demonstration of engineering any arbitrary state by controlled dynamics generated by quantum walk has been reported\,\cite{Gio19}. This strongly supports feasibility of our scheme where a desired probability distribution will act as an additional resource. However, a crucial point of our QRNG scheme is that,  we use the inherent randomness of the quantum dynamics which  is completely controlled under the person or organization who wants to produce random bits for any further cryptographic application. 

QRNG based on state preparation and measurement scheme where state can be manipulated in such a way that the observer of the random sequence of bits can think of it as a perfectly random whereas the output can be completely predictable to the adversary who is sending the states. In our scheme we have shown how the randomness is almost independent of the initial state after few steps of quantum walk and since dynamics involves quantum interference, the output is  random enough to produce random bits string in a secure lab where he/she has the full control of the quantum walk dynamics. The only issue with the hardware failure or malfunctioning, such that dynamics can be highly localized i.e the probability distribution of finding particle within some specific range of positions is really high therefore there might be some chances that it can be predicted but we have been able to produce a solution for that by using an engineered state and dynamics which are capable of producing the uniform or any desired probability distribution, non-occurrence of this desired probability distribution can signal the hardware malfunction therefore the person or organization are ready to rectify it or completely abandon the protocol.

Noise on quantum walk affects the spread of the wavepacket in position space that will proportionally decrease the randomness in the systems. As shown in our results, though a overall randomness does not significantly decrease with noise, contribution of randomness from quantum origin does see a substantial decrease.  This goes well with an established understanding to decrease in entanglement between the coin and position space of the quantum walker with increase in noise.

\vskip 0.2in
\noindent
{\bf Acknowledgment:}

\noindent
CMC would like to thank Department of Science and Technology, Government of India for the Ramanujan Fellowship grant No.:SB/S2/RJN-192/2014. CMC and Anumpam Sarkar would like to thank Shivani Singh and R. Srikanth for useful discussions.

\vskip 0.2in

\noindent  
{\bf Author Contributions:}

\noindent
CMC designed the study, carried out numbercial analysis and prepared figures. AS and CMC together carried out analytical derivation, interpreted the results, and wrote the manuscript.

\vskip 0.2in
\noindent
{\bf Competing Interests :}

\noindent
The authors declare that there are no competing interests.  
  

\newpage

\onecolumngrid
\appendix
\section{ Calculating Randomness of The State After First Step }
\subsection{Coin space}
The initial state of the walker, $ \ket{\psi_{in}}= (\cos\delta\ket{\uparrow}+ e^{i\eta}\sin\delta\ket{\downarrow})\otimes \ket{0}$. The Coin operation we consider
\[
C(\theta)=
  \begin{bmatrix}
    \cos\theta & -i\sin\theta \\
    -i\sin\theta & \cos\theta
  \end{bmatrix}\]
  Therefore, 
  \begin{align*}  
 & \Big[C(\theta)\otimes \mathds{1}\Big]\ket{\psi_{in}}= (\cos\theta\cos\delta-ie^{i\eta}\sin\theta\sin\delta)\ket{\uparrow}+ (e^{i\eta}\cos\theta\sin\delta-i\sin{\theta}\cos{\delta})\ket{\downarrow}\\
& S\Big[C(\theta)\otimes \mathds{1}\Big]\ket{\psi_{in}}=(\cos\theta\cos\delta-ie^{i\eta}\sin\theta\sin\delta)\ket{\uparrow}\otimes \ket{-1}+ (e^{i\eta}\cos\theta\sin\delta-i\sin{\theta}\cos{\delta})\ket{\downarrow}\otimes \ket{1}\\
  \end{align*}
  After one step the state of the walker,
  $$ \ket{\psi_{1}}= S\Big[C(\theta)\otimes \mathds{1}\Big]\ket{\psi_{in}}= (\cos\theta\cos\delta-ie^{i\eta}\sin\theta\sin\delta)\ket{\uparrow}\otimes \ket{-1}+ (e^{i\eta}\cos\theta\sin\delta-i\sin{\theta}\cos{\delta})\ket{\downarrow}\otimes \ket{1}$$
  So the corresponding density matrix $\rho_{1}$ will be,
  \begin{align*}
  &\rho_{1}= \ket{\psi_{1}}\bra{\psi_{1}}\\
  & =\Big[(\cos\theta\cos\delta-ie^{i\eta}\sin\theta\sin\delta)\ket{\uparrow}\otimes \ket{-1}+ (e^{i\eta}\cos\theta\sin\delta-i\sin{\theta}\cos{\delta})\ket{\downarrow}\otimes \ket{1}\Big]\times\\ & \Big[(\cos{\theta}\cos{\delta}+ie^{-i\eta}\sin{\theta}\sin{\delta})\bra{\uparrow}\otimes\bra{-1}+(e^{-i\eta}\cos{\theta}\sin{\delta}+i\sin{\theta}\cos{\delta})\bra{\downarrow}\otimes\bra{1}\Big]\\
 & =\Big(\cos^{2}{\theta}\cos^{2}{}-ie^{i\eta}\sin{\theta}\cos{\theta}\cos{\delta}\sin{\delta}+ie^{-i\eta}\sin{\theta}\cos{\theta}\sin{\delta}\cos{\delta}+\sin^{2}{\theta}\sin^{2}{\delta}\Big)\ket{\uparrow}\bra{\uparrow}\otimes\ket{-1}\bra{-1}\\
 & +\Big(e^{-i\eta}\cos^{2}{\theta}\sin{\delta}\cos{\delta}+i\sin{\theta}\cos{\theta}\cos^{2}{\delta}-i\sin{\theta}\cos{\theta}\sin^{2}{\delta}+ e^{i\eta}\sin^{2}{\theta}\sin{\delta}\cos{\delta}\Big)\ket{\uparrow}\bra{\downarrow}\otimes\ket{-1}\bra{1}\\
 &+ \Big(e^{i\eta}\cos^{2}{\theta}\sin{\delta}\cos{\delta}+i\sin{\theta}\cos{\theta}\sin^{2}{\delta}-i \sin{\theta}\cos{\theta}\cos^{2}{\delta}+e^{-i\eta}\sin^{2}{\theta}\sin{\delta}\cos{\delta}\Big)\ket{\downarrow}\bra{\uparrow}\otimes\ket{1}\bra{-1}\\
 &+\Big(\cos^{2}{\theta}\sin^{2}{\delta}+ie^{i\eta}\sin{\theta}\cos{\theta}\sin{\delta}\cos{\delta}-ie^{-i\eta}\sin{\theta}\cos{\theta}\sin{\delta}\cos{\delta}+\sin^{2}{\theta}\cos^{2}{\delta}\Big)\ket{\downarrow}\bra{\downarrow}\otimes\ket{1}\bra{1}
  \end{align*}
  Now by tracing out the position space we'll get the reduced density matrix of the coin space, denoted by $\rho_{1}^{c}$
  \begin{align*}
 & \rho_{1}^{c}= Tr_{p}(\rho_{1})\\
  & =\Big(\cos^{2}{\theta}\cos^{2}{\delta}-ie^{i\eta}\sin{\theta}\cos{\theta}\cos{\delta}\sin{\delta}+ie^{-i\eta}\sin{\theta}\cos{\theta}\sin{\delta}\cos{\delta}+\sin^{2}{\theta}\sin^{2}{\delta}\Big)\ket{\uparrow}\bra{\uparrow}\\
  &+\Big(\cos^{2}{\theta}\sin^{2}{\delta}+ie^{i\eta}\sin{\theta}\cos{\theta}\sin{\delta}\cos{\delta}-ie^{-i\eta}\sin{\theta}\cos{\theta}\sin{\delta}\cos{\delta}+\sin^{2}{\theta}\cos^{2}{\delta}\Big)\ket{\downarrow}\bra{\downarrow}
  \end{align*}
  By simplifying the expression above we get the probability of obtaining $\ket{\uparrow}$ coin state, following the same pocedure to find the randomness in initial sate
   $$P_{\uparrow}= \Big[(\cos{\theta}\cos{\delta}+\sin{\theta}\sin{\delta})^{2}- 2\sin{\theta}\cos{\theta}\sin{\delta}\cos{\delta}(1-\sin{\eta})\Big]$$
 Similarly for the coin state $\ket{\downarrow}$,
 $$P_{\downarrow}= \Big[(\cos{\theta}\sin{\delta}+\sin{\theta}\cos{\delta})^{2}- 2\sin{\theta}\cos{\theta}\sin{\delta}\cos{\delta}(1+\sin{\eta})\Big]$$
 So the amount of randomness inherited by the state after $1$ step quantum walk, denoted by $R_{i}$,
\begin{align*}
R_{i}(\rho_{1}^{c})= &-(P_{\uparrow}\ln P_{\uparrow}+ P_{\downarrow}\ln P_{\downarrow})\\
&=-\Bigg[\Big[(\cos{\theta}\cos{\delta}+\sin{\theta}\sin{\delta})^{2}- 2\sin{\theta}\cos{\theta}\sin{\delta}\cos{\delta}(1-\sin{\eta})\Big]\\
&\times\ln\Big[(\cos{\theta}\cos{\delta}+\sin{\theta}\sin{\delta})^{2}- 2\sin{\theta}\cos{\theta}\sin{\delta}\cos{\delta}(1-\sin{\eta})\Big]\\
&+\Big[(\cos{\theta}\sin{\delta}+\sin{\theta}\cos{\delta})^{2}- 2\sin{\theta}\cos{\theta}\sin{\delta}\cos{\delta}(1+\sin{\eta})\Big]\\
&\times\ln\Big[(\cos{\theta}\sin{\delta}+\sin{\theta}\cos{\delta})^{2}- 2\sin{\theta}\cos{\theta}\sin{\delta}\cos{\delta}(1+\sin{\eta})\Big]\Bigg]
\end{align*}
\subsection{Position space}
The density matrix after first step
\begin{align*}
  &\rho_{1}= \ket{\psi_{1}}\bra{\psi_{1}}\\
 & =\Big(\cos^{2}{\theta}\cos^{2}{}-ie^{i\eta}\sin{\theta}\cos{\theta}\cos{\delta}\sin{\delta}+ie^{-i\eta}\sin{\theta}\cos{\theta}\sin{\delta}\cos{\delta}+\sin^{2}{\theta}\sin^{2}{\delta}\Big)\ket{\uparrow}\bra{\uparrow}\otimes\ket{-1}\bra{-1}\\
 & +\Big(e^{-i\eta}\cos^{2}{\theta}\sin{\delta}\cos{\delta}+i\sin{\theta}\cos{\theta}\cos^{2}{\delta}-i\sin{\theta}\cos{\theta}\sin^{2}{\delta}+ e^{i\eta}\sin^{2}{\theta}\sin{\delta}\cos{\delta}\Big)\ket{\uparrow}\bra{\downarrow}\otimes\ket{-1}\bra{1}\\
 &+ \Big(e^{i\eta}\cos^{2}{\theta}\sin{\delta}\cos{\delta}+i\sin{\theta}\cos{\theta}\sin^{2}{\delta}-i \sin{\theta}\cos{\theta}\cos^{2}{\delta}+e^{-i\eta}\sin^{2}{\theta}\sin{\delta}\cos{\delta}\Big)\ket{\downarrow}\bra{\uparrow}\otimes\ket{1}\bra{-1}\\
 &+\Big(\cos^{2}{\theta}\sin^{2}{\delta}+ie^{i\eta}\sin{\theta}\cos{\theta}\sin{\delta}\cos{\delta}-ie^{-i\eta}\sin{\theta}\cos{\theta}\sin{\delta}\cos{\delta}+\sin^{2}{\theta}\cos^{2}{\delta}\Big)\ket{\downarrow}\bra{\downarrow}\otimes\ket{1}\bra{1}
  \end{align*}
  Tracing out the coin space resulting position space density matrix would be
  \begin{align*}
  & \rho_{1}^{p}= Tr_{c}(\rho_{1})\\
  & =\Big(\cos^{2}{\theta}\cos^{2}{\delta}-ie^{i\eta}\sin{\theta}\cos{\theta}\cos{\delta}\sin{\delta}+ie^{-i\eta}\sin{\theta}\cos{\theta}\sin{\delta}\cos{\delta}+\sin^{2}{\theta}\sin^{2}{\delta}\Big)\ket{-1}\bra{-1}\\
  &+\Big(\cos^{2}{\theta}\sin^{2}{\delta}+ie^{i\eta}\sin{\theta}\cos{\theta}\sin{\delta}\cos{\delta}-ie^{-i\eta}\sin{\theta}\cos{\theta}\sin{\delta}\cos{\delta}+\sin^{2}{\theta}\cos^{2}{\delta}\Big)\ket{1}\bra{1}
  \end{align*}
  Therefore the amount of randomness contained in the state,
  \begin{align*}
&R_{i}(\rho_{1}^{p})
=-\Bigg[\Big[(\cos{\theta}\cos{\delta}+\sin{\theta}\sin{\delta})^{2}- 2\sin{\theta}\cos{\theta}\sin{\delta}\cos{\delta}(1-\sin{\eta})\Big]\\
&\times\ln\Big[(\cos{\theta}\cos{\delta}+\sin{\theta}\sin{\delta})^{2}- 2\sin{\theta}\cos{\theta}\sin{\delta}\cos{\delta}(1-\sin{\eta})\Big]\\
&+\Big[(\cos{\theta}\sin{\delta}+\sin{\theta}\cos{\delta})^{2}- 2\sin{\theta}\cos{\theta}\sin{\delta}\cos{\delta}(1+\sin{\eta})\Big]\\
&\times\ln\Big[(\cos{\theta}\sin{\delta}+\sin{\theta}\cos{\delta})^{2}- 2\sin{\theta}\cos{\theta}\sin{\delta}\cos{\delta}(1+\sin{\eta})\Big]\Bigg]
\end{align*}
Interesting thing is to notice that after first step, amount of quantified randomness in position space and coin space are the same which is physically reasonable because of the fact that degees of freedom for coin and position space is same i.e two. Therefore the corresponding probability amplitudes for the two degrees of freedom contribute in the randomness as the result randomness corresponding two spaces will be same.
 \section{ Calculating Randomness of The State After Second Step}
\subsection{Coin space}
At the start of the second step the state of the walker is,
$$ \ket{\psi_{1}}= (\cos\theta\cos\delta-ie^{i\eta}\sin\theta\sin\delta)\ket{\uparrow}\otimes \ket{-1}+ (e^{i\eta}\cos\theta\sin\delta-i\sin{\theta}\cos{\delta})\ket{\downarrow}\otimes \ket{1}$$
\begin{align*}
 \Big[C(\theta)\otimes\mathds{1}\Big]\ket{\psi_{1}}= &\Big[(\cos^2{\theta}\cos{\delta}-ie^{i\eta}\sin{\theta}\cos{\theta}\sin{\delta})\ket\uparrow\otimes\ket{-1}+(-i\sin{\theta}\cos{\theta}\cos{\delta}-e^{i\eta}\sin^2{\theta}\sin{\delta})\ket{\downarrow}\otimes\ket{-1}\\
&+(-ie^{i\eta}\sin{\theta}\cos{\theta}\sin{\delta}-\sin^2{\theta}\cos{\delta})\ket{\uparrow}\otimes\ket{1}+(e^{i\eta}\cos^2{\theta}\sin{\delta}-i\sin{\theta}\cos{\theta}\cos{\delta})\ket{\downarrow}\otimes\ket{1}\Big]
\end{align*}
\begin{align*}
S\Big[C(\theta)\otimes\mathds{1}\Big]\ket{\psi_{1}}= \ket{\psi_{2}}=&\Big[(\cos^2{\theta}\cos{\delta}-ie^{i\eta}\sin{\theta}\cos{\theta}\sin{\delta})\ket{\uparrow}\otimes\ket{-2}+(-i\sin{\theta}\cos{\theta}\cos{\delta}-e^{i\eta}\sin^2{\theta}\sin{\delta})\ket{\downarrow}\otimes\ket{0}\\
&+(-ie^{i\eta}\sin{\theta}\cos{\theta}\sin{\delta}-\sin^2{\theta}\cos{\delta})\ket{\uparrow}\otimes\ket{0}+(e^{i\eta}\cos^2{\theta}\sin{\delta}-i\sin{\theta}\cos{\theta}\cos{\delta})\ket{\downarrow}\otimes\ket{2}\Big]
\end{align*}
The density matrix denoted by $\rho_{2}$
 \begin{align*}
 \rho_{2}=\ket{\psi_{2}}\bra{\psi_{2}}=&\Big[(\cos^2{\theta}\cos{\delta}-ie^{i\eta}\sin{\theta}\cos{\theta}\sin{\delta})\ket{\uparrow}\otimes\ket{-2}+(-i\sin{\theta}\cos{\theta}\cos{\delta}-e^{i\eta}\sin^2{\theta}\sin{\delta})\ket{\downarrow}\otimes\ket{0}\\
&+(-ie^{i\eta}\sin{\theta}\cos{\theta}\sin{\delta}-\sin^2{\theta}\cos{\delta})\ket{\uparrow}\otimes\ket{0}+(e^{i\eta}\cos^2{\theta}\sin{\delta}-i\sin{\theta}\cos{\theta}\cos{\delta})\ket{\downarrow}\otimes\ket{2}\Big]\\
&\Big[(\cos^2{\theta}\cos{\delta}+ie^{-i\eta}\sin{\theta}\cos{\theta}\sin{\delta})\bra{\uparrow}\otimes\bra{-2}+(i\sin{\theta}\cos{\theta}\cos{\delta}-e^{-i\eta}\sin^2{\theta}\sin{\delta})\bra{\downarrow}\otimes\bra{0}\\
&+(ie^{-i\eta}\sin{\theta}\cos{\theta}\sin{\delta}-\sin^2{\theta}\cos{\delta})\bra{\uparrow}\otimes\bra{0}+(e^{-i\eta}\cos^2{\theta}\sin{\delta}+i\sin{\theta}\cos{\theta}\cos{\delta})\bra{\downarrow}\otimes\bra{2}\Big]
\end{align*}
{\begin{align*}
&=\Big[(\cos^{4}{\theta}\cos^{2}{\delta}+ie^{-i\eta}\sin{\theta}\cos^{3}{\theta}\sin{\delta}\cos{\delta}-ie^{i\eta}\sin{\theta}\cos^{3}{\theta}\sin{\delta}\cos{\delta}+\sin^{2}{\theta}\cos^{2}{\theta}\sin^2{\delta})\ket{\uparrow}\bra{\uparrow}\otimes\ket{-2}\bra{-2}\\
&+(i\sin{\theta}\cos^{3}{\theta}\cos^2{\delta}-e^{-i\eta}\sin^{2}{\theta}\cos^{2}{\theta}\sin{\delta}\cos{\delta}+e^{i\eta}\sin^{2}{\theta}\cos^{2}{\theta}\sin{\delta}\cos{\delta}+i\sin^{3}{\theta}\cos{\theta}\sin^{2}{\delta})\ket{\uparrow}\bra{\downarrow}\otimes\ket{-2}\bra{0}\\
&+(ie^{-i\eta}\sin{\theta}\cos^{3}{\theta}\cos{\delta}\sin{\delta}-\sin^{2}{\theta}\cos^{2}{\theta}\cos^{2}{\delta}+\sin^{2}{\theta}\cos^{2}{\theta}\sin^{2}{\delta}+ie^{i\eta}\sin^{3}{\theta}\cos{\theta}\sin{\delta}\cos{\delta})\ket{\uparrow}\bra{\uparrow}\otimes\ket{-2}\bra{0}\\
&+(e^{-i\eta}\cos^{4}{\theta}\cos{\delta}\sin{\delta}+i\sin{\theta}\cos^{3}{\theta}\cos^{2}{\delta}-i\sin{\theta}\cos^{3}{\theta}\sin^{2}{\delta}+e^{i\eta}\sin^{2}{\theta}\cos^{2}{\theta}\sin{\delta}\cos{\delta})\ket{\uparrow}\bra{\downarrow}\otimes\ket{-2}\bra{2}\\
&+(-i\sin{\theta}\cos^{3}{\theta}\cos^{2}{\delta}+e^{-i\eta}\sin^{2}{\theta}\cos^{2}{\theta}\sin{\delta}\cos{\delta}-e^{i\eta}\sin^{2}{\theta}\cos^{2}{\theta}\sin{\delta}\cos{\delta}-i\sin^{3}{\theta}\cos{\theta}\sin^{2}{\delta})\ket{\downarrow}\bra{\uparrow}\otimes\ket{0}\bra{-2}\\
&+(\sin^{2}{\theta}\cos^{2}{\theta}\cos^{2}{\delta}+ie^{-i\eta}\sin^{3}{\theta}\cos{\theta}\sin{\delta}\cos{\delta}-ie^{i\eta}\sin^{3}{\theta}\cos{\theta}\sin{\delta}\cos{\delta}+\sin^{4}{\theta}\sin^{2}{\delta})\ket{\downarrow}\bra{\downarrow}\otimes\ket{0}\bra{0}\\
&+(e^{-i\eta}\sin^{2}{\theta}\cos^{2}{\theta}\sin{\delta}\cos{\delta}-i\sin^{3}{\theta}\cos{\theta}\cos^{2}{\delta}-i\sin^{3}{\theta}\cos{\theta}\sin^{2}{\delta}+e^{i\eta}\sin^{4}{\theta}\sin{\delta}\cos{\delta})\ket{\downarrow}\bra{\uparrow}\otimes\ket{0}\bra{0}\\
&+(-ie^{-i\eta}\sin{\theta}\cos^{3}{\theta}\sin{\delta}\cos{\delta}+\sin^{2}{\theta}\cos^{2}{\theta}\cos^{2}{\delta}-\sin^{2}{\theta}\cos^{2}{\theta}\sin^{2}{\delta}-ie^{i\eta}\sin^{3}{\theta}\cos{\theta}\sin{\delta}\cos{\delta})\ket{\downarrow}\bra{\downarrow}\otimes\ket{0}\bra{2}\\
&+(-ie^{i\eta}\sin{\theta}\cos^{3}{\theta}\sin{\delta}\cos{\delta}+\sin^{2}{\theta}\cos^{2}{\theta}\sin^{2}{\delta}-\sin^{2}{\theta}\cos^{2}{\theta}\cos^{2}{\delta}-ie^{-i\eta}\sin^{3}{\theta}\cos{\theta}\sin{\delta}\cos{\delta})\ket{\uparrow}\bra{\uparrow}\otimes\ket{0}\bra{-2}\\
&+(e^{i\eta}\sin^{2}{\theta}\cos^{2}{\theta}\sin{\delta}\cos{\delta}+i\sin^{3}{\theta}\cos{\theta}\sin^{2}{\delta}-i\sin^{3}{\theta}\cos{\theta}\cos^{2}{\delta}+e^{-i\eta}\sin^{4}{\theta}\sin{\delta}\cos{\delta})\ket{\uparrow}\bra{\downarrow}\otimes\ket{0}\bra{0}\\
&+(\sin^{2}{\theta}\cos^{2}{\theta}\sin^{2}{\delta}+ie^{i\eta}\sin^{3}{\theta}\cos{\theta}\sin{\delta}\cos{\delta}-ie^{-i\eta}\sin^{3}{\theta}\cos{\theta}\sin{\delta}\cos{\delta}+\sin^{4}{\theta}\cos^{2}{\delta})\ket{\uparrow}\bra{\uparrow}\otimes\ket{0}\bra{0}\\
&+(-i\sin{\theta}\cos^{3}{\theta}\sin^2{\delta}+e^{i\eta}\sin^{2}{\theta}\cos^{2}{\theta}\sin{\delta}\cos{\delta}-e^{-i\eta}\sin^{2}{\theta}\cos^{2}{\theta}\sin{\delta}\cos{\delta}-i\sin^{3}{\theta}\cos{\theta}\cos^{2}{\delta})\ket{\uparrow}\bra{\downarrow}\otimes\ket{0}\bra{2}\\
&+ (e^{i\eta}\cos^{4}{\theta}\sin{\delta}\cos{\delta}+i\sin{\theta}\cos^{3}{\theta}\sin^{2}{\delta}-i\sin{\theta}\cos^{3}{\theta}\cos^2{\delta}+e^{-i\eta}\sin^{2}{\theta}\cos^{2}{\theta}\sin{\delta}\cos{\delta})\ket{\downarrow}\bra{\uparrow}\otimes\ket{2}\bra{-2}\\
&+(ie^{i\eta}\sin{\theta}\cos^{3}{\theta}\sin{\delta}\cos{\delta}-\sin^{2}{\theta}\cos^{2}{\theta}\sin^{2}{\delta}+\sin^{2}{\theta}\cos^{2}{\theta}\cos^{2}{\delta}+ie^{-i\eta}\sin^{3}{\theta}\cos{\theta}\sin{\delta}\cos{\delta})\ket{\downarrow}\bra{\downarrow}\otimes\ket{2}\bra{0}\\
&+(i\sin{\theta}\cos^{3}{\theta}\sin^{2}{\delta}-e^{i\eta}\sin^{2}{\theta}\cos^{2}{\theta}\sin{\delta}\cos{\delta}+e^{-i\eta}\sin^{2}{\theta}\cos^{2}{\theta}\sin{\delta}\cos{\delta}+i\sin^{3}{\theta}\cos{\theta}\cos^{2}{\delta})\ket{\downarrow}\bra{\uparrow}\otimes\ket{2}\bra{0}\\
&+(\cos^{4}{\theta}\sin^{2}{\delta}+ie^{i\eta}\sin{\theta}\cos^{3}{\theta}\sin{\delta}\cos{\delta}-ie^{-i\eta}\sin{\theta}\cos^{3}{\theta}\sin{\delta}\cos{\delta}+\sin^{2}{\theta}\cos^{2}{\theta}\cos^{2}{\delta})\ket{\downarrow}
\bra{\downarrow}\otimes\ket{2}\bra{2}\Big]
 \end{align*}
Now the reduced density matrix for the coin space
\begin{align*}
\rho_{2}^{c}=&\Big[(\cos^{4}{\theta}\cos^{2}{\delta}+ie^{-i\eta}\sin{\theta}\cos^{3}{\theta}\sin{\delta}\cos{\delta}-ie^{i\eta}\sin{\theta}\cos^{3}{\theta}\sin{\delta}\cos{\delta}+\sin^{2}{\theta}\cos^{2}{\theta}\sin^2{\delta})\ket{\uparrow}\bra{\uparrow}\\
&++(\sin^{2}{\theta}\cos^{2}{\theta}\cos^{2}{\delta}+ie^{-i\eta}\sin^{3}{\theta}\cos{\theta}\sin{\delta}\cos{\delta}-ie^{i\eta}\sin^{3}{\theta}\cos{\theta}\sin{\delta}\cos{\delta}+\sin^{4}{\theta}\sin^{2}{\delta})\ket{\downarrow}\bra{\downarrow}\\
&+(\sin^{2}{\theta}\cos^{2}{\theta}\sin^{2}{\delta}+ie^{i\eta}\sin^{3}{\theta}\cos{\theta}\sin{\delta}\cos{\delta}-ie^{-i\eta}\sin^{3}{\theta}\cos{\theta}\sin{\delta}\cos{\delta}+\sin^{4}{\theta}\cos^{2}{\delta})\ket{\uparrow}\bra{\uparrow}\\
&+(\cos^{4}{\theta}\sin^{2}{\delta}+ie^{i\eta}\sin{\theta}\cos^{3}{\theta}\sin{\delta}\cos{\delta}-ie^{-i\eta}\sin{\theta}\cos^{3}{\theta}\sin{\delta}\cos{\delta}+\sin^{2}{\theta}\cos^{2}{\theta}\cos^{2}{\delta})\ket{\downarrow}
\bra{\downarrow}\Big]
\end{align*}
 By simplifying the above expression, we get the form
 \begin{align*}
 \rho_{2}^{c}= &(\cos^{4}{\theta}\cos^{2}{\delta}+\sin{\eta}\sin{2\theta}\cos{2\theta}\sin{\delta}\cos{\delta}+2\sin^{2}{\theta}\cos^{2}{\theta}\sin^{2}{\delta}+\sin^{4}{\theta}\cos^{2}{\delta})\ket{\uparrow}\bra{\uparrow}\\
 &+(\cos^{4}{\theta}\sin^{2}{\delta}-\sin{\eta}\sin{2\theta}\cos{2\theta}\sin{\delta}\cos{\delta}+2\sin^2{\theta}\cos^{2}{\theta}\cos^{2}{\delta}+\sin^{4}{\theta}\sin^{2}{\delta})\ket{\downarrow}\bra{\downarrow}
 \end{align*}
 Therefore,
 \begin{align*}R_{i}(\rho_{2}^{c})= -\Bigg[\Big[(\cos^{4}{\theta}\cos^{2}{\delta}+\sin{\eta}\sin{2\theta}\cos{2\theta}\sin{\delta}\cos{\delta}+2\sin^{2}{\theta}\cos^{2}{\theta}\sin^{2}{\delta}+\sin^{4}{\theta}\cos^{2}{\delta})\\
 \ln((\cos^{4}{\theta}\cos^{2}{\delta}+\sin{\eta}\sin{2\theta}\cos{2\theta}\sin{\delta}\cos{\delta}+2\sin^{2}{\theta}\cos^{2}{\theta}\sin^{2}{\delta}+\sin^{4}{\theta}\cos^{2}{\delta}\Big]\\
 +\Big[(\cos^{4}{\theta}\sin^{2}{\delta}-\sin{\eta}\sin{2\theta}\cos{2\theta}\sin{\delta}\cos{\delta}+2\sin^2{\theta}\cos^{2}{\theta}\cos^{2}{\delta}+\sin^{4}{\theta}\sin^{2}{\delta})\\
 \ln((\cos^{4}{\theta}\sin^{2}{\delta}-\sin{\eta}\sin{2\theta}\cos{2\theta}\sin{\delta}\cos{\delta}+2\sin^2{\theta}\cos^{2}{\theta}\cos^{2}{\delta}+\sin^{4}{\theta}\sin^{2}{\delta}))\Big]\Bigg]
 \end{align*}
 \subsection{Position space}
 We'll use the expression of the density matrix after second step derived above.
\begin{align*}
&\rho_{2}= \ket{\psi_{2}}\bra{\psi_{2}} \\
&=\Big[(\cos^{4}{\theta}\cos^{2}{\delta}+ie^{-i\eta}\sin{\theta}\cos^{3}{\theta}\sin{\delta}\cos{\delta}-ie^{i\eta}\sin{\theta}\cos^{3}{\theta}\sin{\delta}\cos{\delta}+\sin^{2}{\theta}\cos^{2}{\theta}\sin^2{\delta})\ket{\uparrow}\bra{\uparrow}\otimes\ket{-2}\bra{-2}\\
&+(i\sin{\theta}\cos^{3}{\theta}\cos^2{\delta}-e^{-i\eta}\sin^{2}{\theta}\cos^{2}{\theta}\sin{\delta}\cos{\delta}+e^{i\eta}\sin^{2}{\theta}\cos^{2}{\theta}\sin{\delta}\cos{\delta}+i\sin^{3}{\theta}\cos{\theta}\sin^{2}{\delta})\ket{\uparrow}\bra{\downarrow}\otimes\ket{-2}\bra{0}\\
&+(ie^{-i\eta}\sin{\theta}\cos^{3}{\theta}\cos{\delta}\sin{\delta}-\sin^{2}{\theta}\cos^{2}{\theta}\cos^{2}{\delta}+\sin^{2}{\theta}\cos^{2}{\theta}\sin^{2}{\delta}+ie^{i\eta}\sin^{3}{\theta}\cos{\theta}\sin{\delta}\cos{\delta})\ket{\uparrow}\bra{\uparrow}\otimes\ket{-2}\bra{0}\\
&+(e^{-i\eta}\cos^{4}{\theta}\cos{\delta}\sin{\delta}+i\sin{\theta}\cos^{3}{\theta}\cos^{2}{\delta}-i\sin{\theta}\cos^{3}{\theta}\sin^{2}{\delta}+e^{i\eta}\sin^{2}{\theta}\cos^{2}{\theta}\sin{\delta}\cos{\delta})\ket{\uparrow}\bra{\downarrow}\otimes\ket{-2}\bra{2}\\
&+(-i\sin{\theta}\cos^{3}{\theta}\cos^{2}{\delta}+e^{-i\eta}\sin^{2}{\theta}\cos^{2}{\theta}\sin{\delta}\cos{\delta}-e^{i\eta}\sin^{2}{\theta}\cos^{2}{\theta}\sin{\delta}\cos{\delta}-i\sin^{3}{\theta}\cos{\theta}\sin^{2}{\delta})\ket{\downarrow}\bra{\uparrow}\otimes\ket{0}\bra{-2}\\
&+(\sin^{2}{\theta}\cos^{2}{\theta}\cos^{2}{\delta}+ie^{-i\eta}\sin^{3}{\theta}\cos{\theta}\sin{\delta}\cos{\delta}-ie^{i\eta}\sin^{3}{\theta}\cos{\theta}\sin{\delta}\cos{\delta}+\sin^{4}{\theta}\sin^{2}{\delta})\ket{\downarrow}\bra{\downarrow}\otimes\ket{0}\bra{0}\\
&+(e^{-i\eta}\sin^{2}{\theta}\cos^{2}{\theta}\sin{\delta}\cos{\delta}-i\sin^{3}{\theta}\cos{\theta}\cos^{2}{\delta}-i\sin^{3}{\theta}\cos{\theta}\sin^{2}{\delta}+e^{i\eta}\sin^{4}{\theta}\sin{\delta}\cos{\delta})\ket{\downarrow}\bra{\uparrow}\otimes\ket{0}\bra{0}\\
&+(-ie^{-i\eta}\sin{\theta}\cos^{3}{\theta}\sin{\delta}\cos{\delta}+\sin^{2}{\theta}\cos^{2}{\theta}\cos^{2}{\delta}-\sin^{2}{\theta}\cos^{2}{\theta}\sin^{2}{\delta}-ie^{i\eta}\sin^{3}{\theta}\cos{\theta}\sin{\delta}\cos{\delta})\ket{\downarrow}\bra{\downarrow}\otimes\ket{0}\bra{2}\\
&+(-ie^{i\eta}\sin{\theta}\cos^{3}{\theta}\sin{\delta}\cos{\delta}+\sin^{2}{\theta}\cos^{2}{\theta}\sin^{2}{\delta}-\sin^{2}{\theta}\cos^{2}{\theta}\cos^{2}{\delta}-ie^{-i\eta}\sin^{3}{\theta}\cos{\theta}\sin{\delta}\cos{\delta})\ket{\uparrow}\bra{\uparrow}\otimes\ket{0}\bra{-2}\\
&+(e^{i\eta}\sin^{2}{\theta}\cos^{2}{\theta}\sin{\delta}\cos{\delta}+i\sin^{3}{\theta}\cos{\theta}\sin^{2}{\delta}-i\sin^{3}{\theta}\cos{\theta}\cos^{2}{\delta}+e^{-i\eta}\sin^{4}{\theta}\sin{\delta}\cos{\delta})\ket{\uparrow}\bra{\downarrow}\otimes\ket{0}\bra{0}\\
&+(\sin^{2}{\theta}\cos^{2}{\theta}\sin^{2}{\delta}+ie^{i\eta}\sin^{3}{\theta}\cos{\theta}\sin{\delta}\cos{\delta}-ie^{-i\eta}\sin^{3}{\theta}\cos{\theta}\sin{\delta}\cos{\delta}+\sin^{4}{\theta}\cos^{2}{\delta})\ket{\uparrow}\bra{\uparrow}\otimes\ket{0}\bra{0}\\
&+(-i\sin{\theta}\cos^{3}{\theta}\sin^2{\delta}+e^{i\eta}\sin^{2}{\theta}\cos^{2}{\theta}\sin{\delta}\cos{\delta}-e^{-i\eta}\sin^{2}{\theta}\cos^{2}{\theta}\sin{\delta}\cos{\delta}-i\sin^{3}{\theta}\cos{\theta}\cos^{2}{\delta})\ket{\uparrow}\bra{\downarrow}\otimes\ket{0}\bra{2}\\
&+ (e^{i\eta}\cos^{4}{\theta}\sin{\delta}\cos{\delta}+i\sin{\theta}\cos^{3}{\theta}\sin^{2}{\delta}-i\sin{\theta}\cos^{3}{\theta}\cos^2{\delta}+e^{-i\eta}\sin^{2}{\theta}\cos^{2}{\theta}\sin{\delta}\cos{\delta})\ket{\downarrow}\bra{\uparrow}\otimes\ket{2}\bra{-2}\\
&+(ie^{i\eta}\sin{\theta}\cos^{3}{\theta}\sin{\delta}\cos{\delta}-\sin^{2}{\theta}\cos^{2}{\theta}\sin^{2}{\delta}+\sin^{2}{\theta}\cos^{2}{\theta}\cos^{2}{\delta}+ie^{-i\eta}\sin^{3}{\theta}\cos{\theta}\sin{\delta}\cos{\delta})\ket{\downarrow}\bra{\downarrow}\otimes\ket{2}\bra{0}\\
&+(i\sin{\theta}\cos^{3}{\theta}\sin^{2}{\delta}-e^{i\eta}\sin^{2}{\theta}\cos^{2}{\theta}\sin{\delta}\cos{\delta}+e^{-i\eta}\sin^{2}{\theta}\cos^{2}{\theta}\sin{\delta}\cos{\delta}+i\sin^{3}{\theta}\cos{\theta}\cos^{2}{\delta})\ket{\downarrow}\bra{\uparrow}\otimes\ket{2}\bra{0}\\
&+(\cos^{4}{\theta}\sin^{2}{\delta}+ie^{i\eta}\sin{\theta}\cos^{3}{\theta}\sin{\delta}\cos{\delta}-ie^{-i\eta}\sin{\theta}\cos^{3}{\theta}\sin{\delta}\cos{\delta}+\sin^{2}{\theta}\cos^{2}{\theta}\cos^{2}{\delta})\ket{\downarrow}
\bra{\downarrow}\otimes\ket{2}\bra{2}\Big]
 \end{align*}
 Now reduced density matrix of position space will be,
 \begin{align*}
 &\rho_{2}^{p}= \Big[(\cos^{4}{\theta}\cos^{2}{\delta}+ie^{-i\eta}\sin{\theta}\cos^{3}{\theta}\sin{\delta}\cos{\delta}-ie^{i\eta}\sin{\theta}\cos^{3}{\theta}\sin{\delta}\cos{\delta}+\sin^{2}{\theta}\cos^{2}{\theta}\sin^2{\delta})\ket{-2}\bra{-2}\\
 &+(ie^{-i\eta}\sin{\theta}\cos^{3}{\theta}\cos{\delta}\sin{\delta}-\sin^{2}{\theta}\cos^{2}{\theta}\cos^{2}{\delta}+\sin^{2}{\theta}\cos^{2}{\theta}\sin^{2}{\delta}+ie^{i\eta}\sin^{3}{\theta}\cos{\theta}\sin{\delta}\cos{\delta})\ket{-2}\bra{0}\\
 &+(\sin^{2}{\theta}\cos^{2}{\theta}\cos^{2}{\delta}+ie^{-i\eta}\sin^{3}{\theta}\cos{\theta}\sin{\delta}\cos{\delta}-ie^{i\eta}\sin^{3}{\theta}\cos{\theta}\sin{\delta}\cos{\delta}+\sin^{4}{\theta}\sin^{2}{\delta})\ket{0}\bra{0}\\
 &+(-ie^{-i\eta}\sin{\theta}\cos^{3}{\theta}\sin{\delta}\cos{\delta}+\sin^{2}{\theta}\cos^{2}{\theta}\cos^{2}{\delta}-\sin^{2}{\theta}\cos^{2}{\theta}\sin^{2}{\delta}-ie^{i\eta}\sin^{3}{\theta}\cos{\theta}\sin{\delta}\cos{\delta})\ket{0}\bra{2}\\
 &+(-ie^{i\eta}\sin{\theta}\cos^{3}{\theta}\sin{\delta}\cos{\delta}+\sin^{2}{\theta}\cos^{2}{\theta}\sin^{2}{\delta}-\sin^{2}{\theta}\cos^{2}{\theta}\cos^{2}{\delta}-ie^{-i\eta}\sin^{3}{\theta}\cos{\theta}\sin{\delta}\cos{\delta})\ket{0}\bra{-2}\\
 &+(\sin^{2}{\theta}\cos^{2}{\theta}\sin^{2}{\delta}+ie^{i\eta}\sin^{3}{\theta}\cos{\theta}\sin{\delta}\cos{\delta}-ie^{-i\eta}\sin^{3}{\theta}\cos{\theta}\sin{\delta}\cos{\delta}+\sin^{4}{\theta}\cos^{2}{\delta})\ket{0}\bra{0}\\
 &+(ie^{i\eta}\sin{\theta}\cos^{3}{\theta}\sin{\delta}\cos{\delta}-\sin^{2}{\theta}\cos^{2}{\theta}\sin^{2}{\delta}+\sin^{2}{\theta}\cos^{2}{\theta}\cos^{2}{\delta}+ie^{-i\eta}\sin^{3}{\theta}\cos{\theta}\sin{\delta}\cos{\delta})\ket{2}\bra{0}\\
 &+(\cos^{4}{\theta}\sin^{2}{\delta}+ie^{i\eta}\sin{\theta}\cos^{3}{\theta}\sin{\delta}\cos{\delta}-ie^{-i\eta}\sin{\theta}\cos^{3}{\theta}\sin{\delta}\cos{\delta}+\sin^{2}{\theta}\cos^{2}{\theta}\cos^{2}{\delta})\ket{2}\bra{2}\Big]
 \end{align*}
 Only diagonal terms in a given basis will contribute to the randomness and by simplifying the above expression we quantify the randomness corresponding position space as,
 \begin{align*}
 R_{i}(\rho_{2}^{p})= &-\Bigg[(\cos^{4}{\theta}\cos^{2}{\delta}+ie^{-i\eta}\sin{\theta}\cos^{3}{\theta}\sin{\delta}\cos{\delta}-ie^{i\eta}\sin{\theta}\cos^{3}{\theta}\sin{\delta}\cos{\delta}+\sin^{2}{\theta}\cos^{2}{\theta}\sin^2{\delta})\\
 &\times\ln(\cos^{4}{\theta}\cos^{2}{\delta}+ie^{-i\eta}\sin{\theta}\cos^{3}{\theta}\sin{\delta}\cos{\delta}-ie^{i\eta}\sin{\theta}\cos^{3}{\theta}\sin{\delta}\cos{\delta}+\sin^{2}{\theta}\cos^{2}{\theta}\sin^2{\delta})\\
&+\sin^{2}{\theta}\ln(\sin^{2}{\theta})+(\cos^{4}{\theta}\sin^{2}{\delta}+ie^{i\eta}\sin{\theta}\cos^{3}{\theta}\sin{\delta}\cos{\delta}-ie^{-i\eta}\sin{\theta}\cos^{3}{\theta}\sin{\delta}\cos{\delta}+\sin^{2}{\theta}\cos^{2}{\theta}\cos^{2}{\delta})\\
&\times \ln(\cos^{4}{\theta}\sin^{2}{\delta}+ie^{i\eta}\sin{\theta}\cos^{3}{\theta}\sin{\delta}\cos{\delta}-ie^{-i\eta}\sin{\theta}\cos^{3}{\theta}\sin{\delta}\cos{\delta}+\sin^{2}{\theta}\cos^{2}{\theta}\cos^{2}{\delta})\Bigg]
 \end{align*}
 Here an extra term is coming compared to coin space randomness expression because after $2$-nd step in position space particle has three degrees of freedom in SQW scenario but for coin space it's being two. From here we can easily witness the advantage of using position space and benefit of using it with more number of steps.

\end{document}